\documentclass[aps,prl,superscriptaddress,showpacs,twocolumn]{revtex4}
\usepackage{graphicx}
\usepackage{color}
\usepackage{amssymb}

\begin{document}

\affiliation{Institute of Physics, P.O.B.304, HR-10 000, Zagreb, Croatia}

\title{Two component butterfly hysteresis in Ru1222 ruthenocuprate}

\author{I. \v{Z}ivkovi\'{c}, D. Drobac, M. Prester}

\affiliation{Institute of Physics, Zagreb, Croatia}

\date{\today}
\begin{abstract}
We report detailed studies of the ac susceptibility butterfly hysteresis on the Ru1222 ruthenocuprate compounds. Two separate contributions to these hysteresis have been identified and studied. One contribution is ferromagnetic-like and is characterized by the coercive field maximum. Another contribution, represented by the so called inverted maximum, is related to the unusual inverted loops, unique feature of Ru1222 butterfly hysteresis. The different nature of the two identified magnetic contributions is proved by the different temperature dependences involved. By lowering the temperature the inverted peak gradually disappears while the coercive field slowly raises. If the maximum dc field for the hysteresis is increased, the size of the inverted part of the butterfly hysteresis monotonously grows while the position of the peak saturates. In reaching saturation exponential field dependence has been demonstrated to take place. At T = 78 K the saturation field is 42 Oe.
\end{abstract}

\maketitle

\section{Introduction}

The interplay between superconducting and magnetic order in ruthenocuprates raised considerable interest for these materials. They exist in three modifications, RuSr$_{2}$RECu$_{2}$O$_{8}$ (Ru1212), RuSr$_{2}$RE$_{2-x}$Ce$_{x}$Cu$_{2}$O$_{10}$ (Ru1222) and, recently synthesized~\cite{awana2005}, RuSr$_{2}$RECe$_{2}$Cu$_{2}$O$_{12}$ (Ru1232) (RE = Gd, Eu for Ru1212 and Ru1222; RE = Y, Dy for Ru1232). Neutron scattering measurements on Ru1212~\cite{lynn2000} have revealed the existence of the antiferromagnetic order in Ru-sublattice below 130 K, although ferromagnetic-like features have been observed in magnetization measurements~\cite{williams2000}. Measurements of microwave absorption~\cite{pozek2001}, ac susceptibility~\cite{zivkovic2002a}, NMR~\cite{sakai2003} have shown an evidence of the spontaneous vortex phase as an explanation for the coexistence of the superconductivity and magnetism below SC transition around 30 K. On the other hand, Xue et al.~\cite{xue2003a} have suggested the nanoscale separation between ferromagnetic and antiferromagnetic islands.

As for the Ru1222 composition, the magnetic structure is still unknown. ESR measurements~\cite{yoshida2003} have displayed a ferromagnetic resonance below T$_{M1}$ = 180 K with an enhancement in magnetism around T$_{M2}$ = 100 K. Xue et al.~\cite{xue2003b} have found evidence of the clusters above the main peak at T$_{M2}$ (around 100 K, depends on the Eu/Ce ratio). Detailed magnetization and M\"ossbauer study~\cite{felner2004-5} have shown two component magnetism which has been also supported with the muon spin rotation results~\cite{shengelaya2004}. Spin-glass-like behavior, as suggested by Cardoso et al.~\cite{cardoso2003}, occurs at T$_{M2}$. Various dynamical features have been reported by us~\cite{zivkovic2002b}, including pronounced time relaxation of ac susceptibility and the inverted butterfly hysteresis loops in ac susceptibility.

In this paper we present further investigation of the unusual behavior of the Ru1222 material. We have concentrated on the peculiarities of the butterfly hysteresis and have found supporting evidence in favor of the above mentioned two component magnetism picture.

\section{Experimental details}

Samples used in this study are the same as the ones used in our previous article~\cite{zivkovic2002b}. Ac susceptibility measurements were taken by the use of commercial CryoBIND system with the frequency of the alternating field equal to 431 Hz and the field amplitude of 0.1 Oe. The stabilization of the temperature for the hysteresis measurement was better than 10 mK.

\section{Experimental results}

\begin{figure}[t!]
\begin{center}
\includegraphics{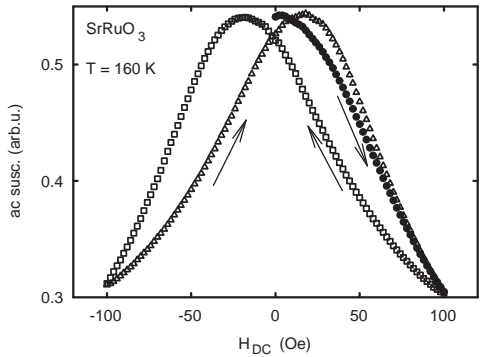}
\caption{Butterfly hysteresis for itinerant ferromagnet SrRuO$_{3}$. ($\bullet$ ) - virgin branch; ($\square$ ) - descending field branch; ($\vartriangle$ ) - ascending field branch.} \label{srruo3hys}
\end{center}
\end{figure}

When a standard ferromagnet, characterized with the domain structure, is swept in a dc field up and down, its magnetization M shows a well-known M--H hysteresis. The characteristic elements of the M--H hysteresis loop are the coercive field and the remanent magnetization. Another type of hysteretic behavior of a ferromagnet can be studied in the mutual inductance arrangement of ac susceptibility technique, imposing an ac field superimposed over the sweeping and cycling dc field. In the latter case one measures a butterfly hysteresis~\cite{salas1993}. In increasing dc field it is characterized by a monotonously decreasing virgin curve, followed by the repetitive branches of further reduced susceptibility values. The repetitive branches are characterized by the characteristic maxima which are related to the coercive field. The values of the coercive field, as obtained from M--H and butterfly hysteresis are not necessarily the same due to the inherent difference between dc and ac techniques. In Fig.~\ref{srruo3hys} butterfly hysteresis is shown for SrRuO$_{3}$, an itinerant ferromagnet.

\begin{figure}[t!]
\begin{center}
\includegraphics{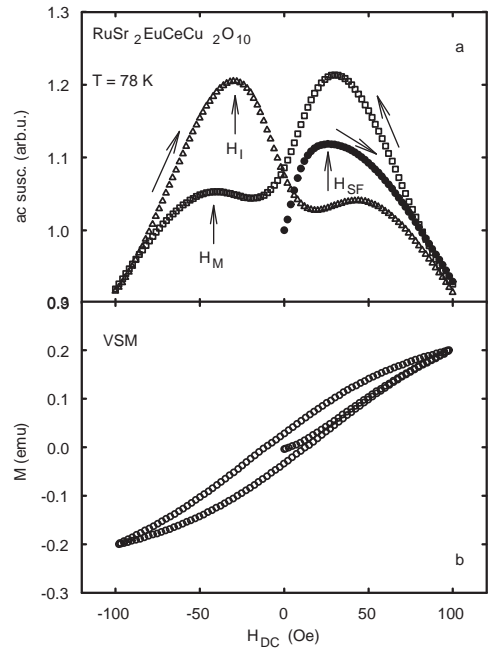}
\caption{(a) Butterfly hysteresis for a ruthenocuprate Ru1222EuCe (x = 1.0) material. ($\bullet$ ) - virgin branch; ($\square$ ) - descending field branch; ($\vartriangle$ ) - ascending field branch. (b) Normal hysteresis obtained with a vibrating sample magnetometer taken at the same temperature and the same maximum dc field as in (a).} \label{ru1222hys}
\end{center}
\end{figure}

At variance with a standard butterfly hysteresis, the one discovered to characterize Ru1222 material~\cite{zivkovic2002b} shows very different behavior when subjected to the sweeping dc field. In Fig.~\ref{ru1222hys}a we present the butterfly hysteresis for the Ru1222EuCe (x = 1.0) material. The virgin branch has a maximum (denoted with H$_{SF}$), followed by the descending field branch which has systematically {\em higher} susceptibility than the virgin branch. Instead of one there are two maxima types, at H$_{I}$ before and H$_{M}$ after H$_{DC}$ = 0. In the ascending field branch, H$_{I}$ maximum appears again, followed by the H$_{M}$ maximum for H$_{DC} > 0$. Such an unusual magnetism hasn't been reported so far and it reveals that the Ru1222 material is a rather unique magnetic system, very different from an ordinary ferromagnet. Fig.~\ref{ru1222hys}b displays the M--H curve measured with a vibrating sample magnetometer (see also~\cite{felner2004-5}). Apart from a somewhat unusual virgin curve (overlapping with the right-hand side hysteresis branch), the M--H hysteresis appears quite regular and standard, implying that the unusual butterfly-hysteresis features rely on magneto-dynamics of the Ru1222 compound.  

\begin{figure}[t!]
\begin{center}
\includegraphics{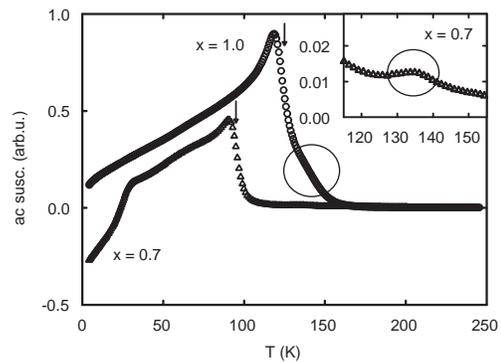}
\caption{Temperature dependence of ac susceptibility, for the two members of the Ru1222 family. Arrows indicate the temperatures where the butterfly hysteresis sets in. Inset shows an enlarged window for the x = 0.7 sample around anomaly at 130 K.} \label{tempdep}
\end{center}
\end{figure}

Temperature dependence of ac susceptibility of the Ru1222 materials is shown in Fig.~\ref{tempdep}. For x = 1.0 the main peak (T$_{M2}$) is at 120 K, while for x = 0.7 it is at 90 K. The x = 0.7 sample is superconducting below 30 K. The anomaly at 130 K can also be seen, especially for the x = 1.0 sample. It has been shown by Felner et al.~\cite{felner2004-5} that this anomaly has a different magnetic origin than the main peak at T$_{M2}$. Also, $\mu $SR experiments~\cite{shengelaya2004} revealed no bulk character of the latter anomaly. For the x = 1.0 composition the main peak at T$_{M2}$ and the anomaly at 130 K are very close and overlapping, so we will use the x = 0.7 material to exclude the possibility that the butterfly hysteresis are connected to this anomaly.

\begin{figure}[t!]
\begin{center}
\includegraphics{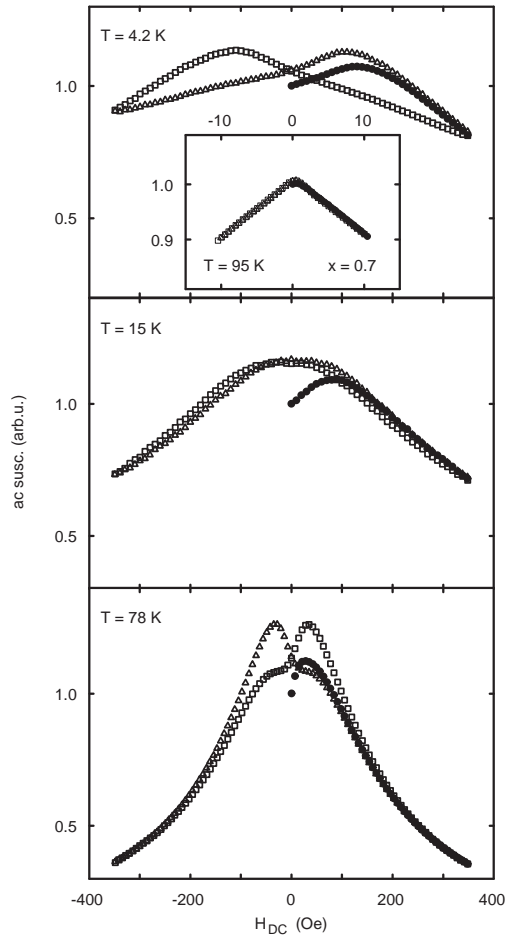}
\caption{Butterfly hysteresis for Ru1222EuCe (x = 1.0) at different temperatures. Inset to top panel: The shape of the hysteresis for Ru1222Eu$_{1.3}$Ce$_{0.7}$ (x = 0.7) at 95 K, just above the temperature at which the inverted peak emerges. This featureless curve resembles the shape of the butterfly hysteresis characterizing Ru1212 material below its magnetic ordering transition~\cite{zivkovic2002a}. At 78 K (bottom panel) the inverted peak is well-defined and dominates in the hysteretic behavior. ($\bullet$ ) - virgin branch; ($\square$ ) - descending field branch; ($\vartriangle$ ) - ascending field branch.} \label{hystemp}
\end{center}
\end{figure}

A qualitative change in the behavior of the butterfly hysteresis can be seen in Fig.~\ref{hystemp}. Three characteristic temperatures have been chosen in order to illustrate the range of peculiar behaviors of butterfly hysteresis covered by varying the temperature. All the curves have been scaled by the value $\chi $(H$_{DC}^{virgin}$ = 0 Oe). The most important feature to notice is the disappearance of the inverted part, characterized by the maximum at H$_{I}$ in Fig.~\ref{ru1222hys}. For temperatures above 30 K the butterfly hysteresis is inverted but below it starts to accommodate the single-peak shape associated with the ferromagnets (Fig.~\ref{srruo3hys}), except for the virgin branch. The H$_{SF}$ maximum in the virgin curve has been linked with the possible spin-flop mechanism~\cite{zivkovic2002a} where the initial AFM state changes into a FM state under the influence of the external dc field.

The exact temperature where the H$_{SF}$ peak and the butterfly hysteresis set in is indicated with the arrows in the Fig~\ref{tempdep}. For the x = 1.0 composition there is just a small difference in the temperature between the main peak at T$_{M2}$ and the anomaly at 130 K and one cannot be sure with which peak to associate this event. For that reason we have measured the x = 0.7 material for which the main peak is at 90 K and it clearly indicates that the observed dynamics is not connected with the small anomaly around 130 K but rather with the main peak at T$_{M2}$ which, in turn, depends on the Eu/Ce ratio.

\begin{figure}[t!]
\begin{center}
\includegraphics{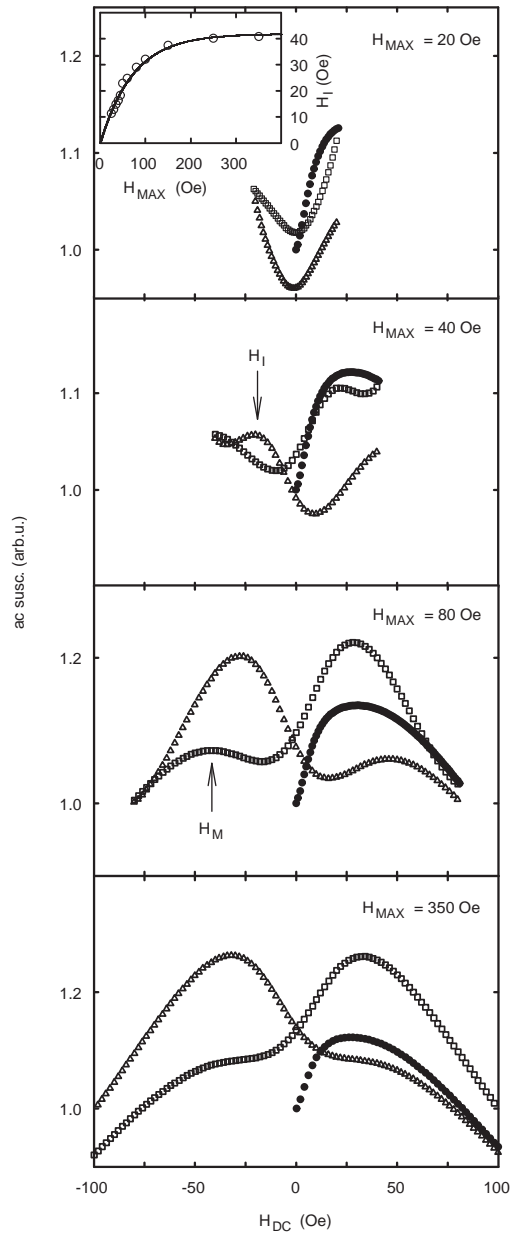}
\caption{Emergence of the inverted part of the butterfly hysteresis for Ru1222EuCe (x = 1.0) sample at T = 78 K as the maximum dc filed is increased. Note how the inverted part stretches above the ferromagnetic background and hides H$_{M}$ for H$_{MAX}$ = 350 Oe. Inset: the maximum of the inverted peak H$_I$ vs. the maximum dc field H$_{MAX}$. The line is the fit to the exponential saturation (see text). ($\bullet$ ) - virgin branch; ($\square$ ) - descending field branch; ($\vartriangle$ ) - ascending field branch.} \label{maxDC}
\end{center}
\end{figure}

In addition to the temperature dependence of the butterfly hysteresis for the Ru1222 material, we have investigated the influence of the maximum dc field one reaches in cycling the dc field. Fig.~\ref{maxDC} shows how the inverted peak (H$_{I}$) emerges for small dc fields (above 25 Oe) and eventually completely overwhelms the underlying ferromagnetic behavior for larger dc fields where it tends to saturate (here {\em larger} denotes a few hundred Oe). After the careful analysis has been done and the ferromagnetic background has been subtracted, one gets the dependence of the inverted peak H$_{I}$ vs. maximum dc field imposed on the system H$_{MAX}$, shown in the inset of Fig.~\ref{maxDC}. The line represents a fit to the formula:
$$H_{I} = H^{\infty }\left( 1 - e^{- \frac{H_{MAX}}{H_C}}\right)$$
with the parameters H$^{ \infty}$ = 42(1) Oe and H$_C$ = 72(4) Oe (correlation coefficient: r = 0.987). At the same time, there is no observable shift in the H$_{M}$ peak, as is expected for a coercive maximum. One should note that there is no sign of the H$_{I}$ peak in the descending field branch (squares) for H$_{DC} <$ 0 nor in the ascending field branch (triangles) for H$_{DC} >$ 0. Also, it seems that the H$_{SF}$ peak is linked only to the virgin curve resembling the S-shaped VSM virgin curve (see Fig.~\ref{srruo3hys}b) and has no influence on the behavior of the inverted part.

\section{Discussion}

As has been recently proposed~\cite{xue2003b,felner2004-5,shengelaya2004}, Ru1222 materials consist of two phases, the minority one that orders around 180 K and the majority one that orders around 100 K (depending on the Eu/Ce ratio). Exact nature of both orderings is still unclear. Although ferromagnetic-like features have been observed, detailed investigation of butterfly hysteresis in Ru1222 reveals a more complex magnetism characterizing this material. Temperature dependence of the inverted part (Fig.~\ref{hystemp}) shows that it emerges at the main peak T$_{M2}$ where the majority phase orders, progressively freezing out as the temperature is lowered. We have investigated samples with x from 0.5 to 1.0 and they all show the same behavior, suggesting similar magnetic ordering to take place in all compositions. The H$_{M}$ peak, which marks the coercive field, gradually increases as the temperature is decreased and reaches the value of 100 Oe for 4.2 K. It can be compared with the values of 250 Oe obtained in~\cite{felner2004-5} for much larger maximum fields (50 kOe). From the Figs.~\ref{hystemp} and~\ref{maxDC} it is obvious that H$_{M}$ is not affected by the presence of the inverted peak, suggesting two separate contributions to the ac susceptibility.

Presently, we are unable to provide a full interpretation of the exponential dependence shown in the inset of the Fig.~\ref{maxDC}. We note however that it indicates the presence of some fundamental interaction which is susceptible to small magnetic fields. Also, a remarkable fact is that no feature can be seen in the magnetization measurements (either VSM or SQUID) that would correspond to the inverted part of the butterfly hysteresis. We suggest that the inverted behavior might reflect the interaction of the two magnetic phases, namely the ferromagnetic clusters and the background matrix, ordering at T$_{M1}$ and at T$_{M2}$, respectively. As suggested by Cardoso et al.~\cite{cardoso2003} the phenomenology of spin glasses could be relevant for Ru1222 as well. Indeed, we have observed the frequency dependence of the main peak at T$_{M2}$(not shown). The magnitude of the shift follows the spin-glass-like behavior~\cite{mydosh} but the magnitude of the ac susceptibility signal is orders of magnitude larger than for the usual spin-glass material. We speculate that the ferromagnetic clusters, randomly distributed and oriented in matrix, could impose a frustration on the surrounding matrix, giving rise to the spin-glass behavior and the observed frequency dependence.

\section{Conclusion}

We have measured the butterfly hysteresis for the Ru1222 material and have found that it consists of two components. The first one comes from the already observed ferromagnetic-like behavior of the compound and is represented by the coercive field peak. The second component is responsible for the onset of the inverted peak in butterfly hysteresis, formed after the dc field reaches its maximum value and starts reducing. The inverted part disappears as the temperature is lowered while the coercive maximum just shifts to larger values. Exponential dependence of the inverted peak vs. maximum dc field has been observed but it's microscopic background is not well understood as yet. Interaction between ferromagnetic clusters and the ordered matrix embedding the clusters has been proposed to account for the observed behavior.

\section{Acknowledgments}

We thank prof. I. Felner for providing us with samples and giving valuable comments.



\end{document}